\documentclass[10pt,conference]{IEEEtran}
\IEEEoverridecommandlockouts
\usepackage{cite}
\usepackage{amsmath,amssymb,amsfonts}
\usepackage{algorithmic}
\usepackage{graphicx}
\usepackage{textcomp}
\usepackage{xcolor}
\usepackage[hyphens]{url} 
\def\BibTeX{{\rm B\kern-.05em{\sc i\kern-.025em b}\kern-.08em
    T\kern-.1667em\lower.7ex\hbox{E}\kern-.125emX}}
\begin{document}

\IEEEoverridecommandlockouts\IEEEpubid{\makebox[\columnwidth]{ 978-1-6654-8045-1/22/\$31.00~\copyright~2022 IEEE \hfill} \hspace{\columnsep}\makebox[\columnwidth]{ }}

\title{Metadata Caching in Presto: Towards Fast Data Processing\\
}

\author{\IEEEauthorblockN{Beinan Wang\IEEEauthorrefmark{1}, Chunxu Tang\IEEEauthorrefmark{2}, Rongrong Zhong\IEEEauthorrefmark{3}, Bin Fan\IEEEauthorrefmark{1}, Yi Wang\IEEEauthorrefmark{1}, Jasmine Wang\IEEEauthorrefmark{1}, Shouwei Chen\IEEEauthorrefmark{1},\\Bowen Ding\IEEEauthorrefmark{1}, Lu Zhang\IEEEauthorrefmark{1}}
\IEEEauthorblockA{
\IEEEauthorrefmark{1}\textit{Alluxio, Inc.}, San Mateo, California, USA\\
\IEEEauthorrefmark{2}\textit{Twitter, Inc.}, San Francisco, California, USA\\
\IEEEauthorrefmark{3}\textit{Celonis, Inc.}, New York City, New York, USA
\\
beinan@alluxio.com, chunxutang@gmail.com, r.zhong@celonis.com, binfan@alluxio.com, hope.wang@alluxio.com, \\
jasmine@alluxio.com, shouwei@alluxio.com, bowen@alluxio.com, lu.zhang@alluxio.com}
}

 \maketitle

\begin{abstract}
    Presto is an open-source distributed SQL query engine for OLAP, aiming for ``SQL on everything''. Since open-sourced in 2013, Presto has been consistently gaining popularity in large-scale data analytics and attracting adoption from a wide range of enterprises. From the development and operation of Presto, we witnessed a significant amount of CPU consumption on parsing column-oriented data files in Presto worker nodes. This blocks some companies, including Meta, from increasing analytical data volumes. 
    
    In this paper, we present a metadata caching layer, built on top of the Alluxio SDK cache and incorporated in each Presto worker node, to cache the intermediate results in file parsing. The metadata cache provides two caching methods: caching the decompressed metadata bytes from raw data files and caching the deserialized metadata objects. Our evaluation of the TPC-DS benchmark on Presto demonstrates that when the cache is warm, the first method can reduce the query's CPU consumption by 10\%-20\%, whereas the second method can minimize the CPU usage by 20\%-40\%.
\end{abstract}


\section{Introduction}
\label{sec.intro}

Presto \cite{sethi2019presto,luo2022batch} is an open-source distributed SQL query engine for OLAP (online analytical processing), targeting ``SQL on everything''. Initially developed by Meta (formerly known as Facebook), Presto has now been widely embraced by a high number of companies, such as Uber \cite{uber-presto}, Twitter \cite{tang2021hybrid,tang2022serving}, and Pinterest \cite{pinterest-presto}. Thanks to its high adaptivity, flexibility, extensibility, and performance, Presto powers a wide spectrum of use cases, ranging from interactive queries (whose latencies are from seconds to minutes), and batch ETL (Extract, Transform, and Load) jobs, to A/B testing. In the past few years, the Presto community has made many critical improvements to the Presto ecosystem including improved readers for nested data processing \cite{presto-parquet}, geospatial analytics \cite{luo2022batch}, file list and footer cache with Alluxio \cite{raptorx}, query cost prediction \cite{tang2021forecasting}, Presto in the cloud \cite{presto-cloud}, etc. 

During the development and support of Presto, we discovered that under the high data volume scenario, the current design can cause an incredibly high CPU consumption on parsing data files, which becomes a performance bottleneck. For example, Meta engineers reported \footnote{https://github.com/prestodb/presto/issues/16337} that they were seeing 30\% CPU resources being used to parse data files, consisting of around three thousand columns. They will need to set aside around 20 GB of memory to hold the 100 million objects if they choose to blindly cache them all. This issue dramatically hinders Meta's adoption of a larger-scale Presto system.

To solve the problem, taking into account of different access frequencies and data volumes between data and metadata, we propose a metadata caching layer based on the current hierarchical cache design in Presto. In Presto, worker nodes are responsible for interacting with external storage and parsing data files. Implemented in the Presto worker, this unified cache layer sits on top of the concrete file format parsing and provides two caching methods: caching the decompressed metadata bytes and caching the deserialized metadata objects. It is aware of the file formats parsed and employs optimization techniques based on concrete formats, including Apache ORC \cite{apache-orc} and Apache Parquet \cite{vohra2016apache}. Our evaluation results show that this metadata caching layer can significantly enhance the performance of Presto.

This paper makes the following contributions:

\begin{enumerate}
    \item Motivated by enterprise-grade practical challenges in operating Presto, we introduce a metadata caching layer based on the Alluxio SDK cache to cache metadata information for parsing data files.
    \item We evaluate the proposed approach on the TPC-DS benchmark and implement a quantitative analysis on the performance tradeoff of reads and writes between caching decompressed metadata bytes and caching deserialized metadata objects.
\end{enumerate}

The remainder of this paper is organized as follows. Section \ref{sec.presto} introduces the overall design of Presto and some implementation details in parsing data files. Section \ref{sec.solution} explains the high-level design of the metadata caching layer and details of the two caching methods. We implement a quantitative analysis of the two caching methods, compared with a no-cache method, in Section \ref{sec.evaluation}. Section \ref{sec.related-work} outlines some related work. Section \ref{sec.conclusion} concludes the paper.

\section{How Presto Works}
\label{sec.presto}

\begin{figure}[thp]
    \centerline{\includegraphics[width=0.4\textwidth]{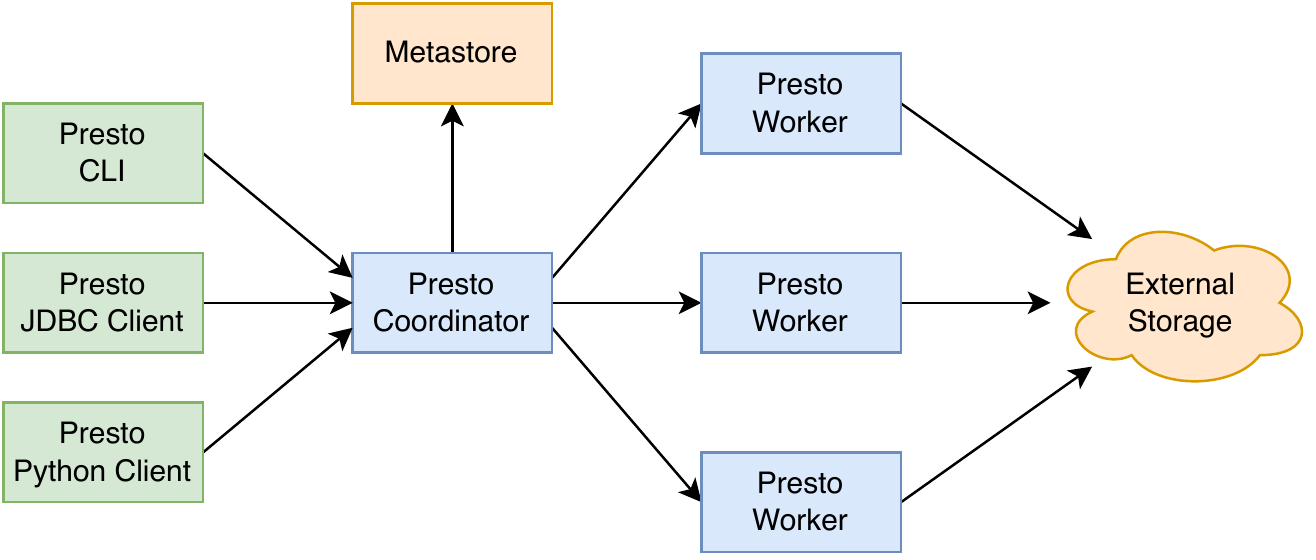}}
    \caption{Architectural design of Presto.}
    \label{fig.presto-arch}
\end{figure}

A typical Presto cluster contains a coordinator node and multiple worker nodes, as shown in Figure \ref{fig.presto-arch}. Presto clients, such as Presto CLI, JDBC client, and Python client, send queries to the coordinator node. The coordinator node takes charge of parsing the queries into abstract syntax trees. Afterward, it plans and optimizes the queries, to generate a distributed execution plan. The coordinator node distributes the plan to the worker nodes, who take charge of the subsequential query processing.

\begin{figure}[t]
    \centerline{\includegraphics[width=0.48\textwidth]{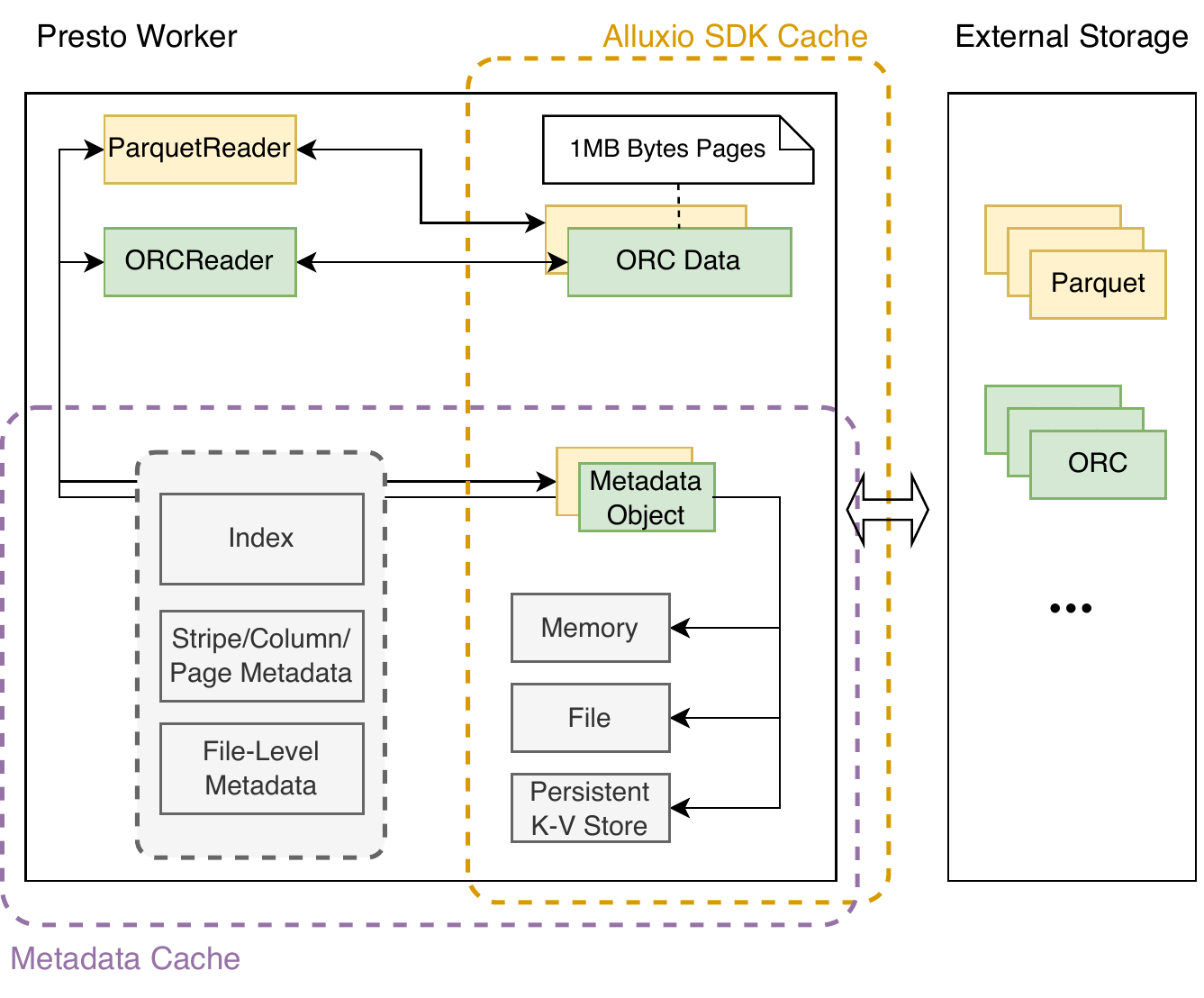}}
    \caption{High-level infrastructure design for the metadata caching and related components.}
    \label{fig.architecture}
\end{figure}

Worker nodes fetch data from external storage systems, such as HDFS (Hadoop Distributed File System), Amazon S3, and GCS (Google Cloud Bucket), and process splits arranged by the coordinator. They open files, parse files, read footers, and read data blocks. The current Presto also supports a local cache in each worker node to improve query performance, but only for data blocks. 

Presto targets OLAP scenarios, hence it supports multiple column-oriented file formats, such as ORC and Parquet. These file formats organize data into columns, where data in the same column is intended to be stored closely. Columnar file formats also usually harness groups of rows (stripes in ORC, row groups in Parquet) to split the data for parallel processing. Row groups' indexes also help Presto skip unnecessary rows when the predicate push-down strategy is enabled. As we found, raw data file parsing becomes a performance bottleneck under a high file number scenario. To mitigate the performance issue, we present a metadata cache in each Presto worker node to store intermediate file parsing metadata.

\section{Metadata Caching}
\label{sec.solution}

\subsection{High-Level Design}

Figure \ref{fig.architecture} depicts the architectural design of the metadata caching and related components. Presto has corresponding readers for file formats. For example, the \textit{ParquetReader} is used to parse and read Parquet raw data files, which can be cached in the Alluxio SDK cache, sitting on top of external storage systems. As raw files are compressed, the readers first decompress the files and then extract metadata information and data. We support two caching methods in the design: caching decompressed metadata bytes and caching deserialized metadata objects, depending on the specific step in the Presto data parsing procedure. The cached metadata information contains metadata for stripes, columns, or pages, indexes, and file-level metadata. Our design supports caching the objects in memory, files, and persistent key-value stores like RocksDB \cite{cao2020characterizing}. We implement configurable cache eviction strategies including FIFO (First-In-First-Out), LRU (Least Recently Used), and LFU (Least Frequently Used).

\subsection{Method I - Caching Decompressed Metadata Bytes}

\begin{figure}[thp]
    \centerline{\includegraphics[width=0.45\textwidth]{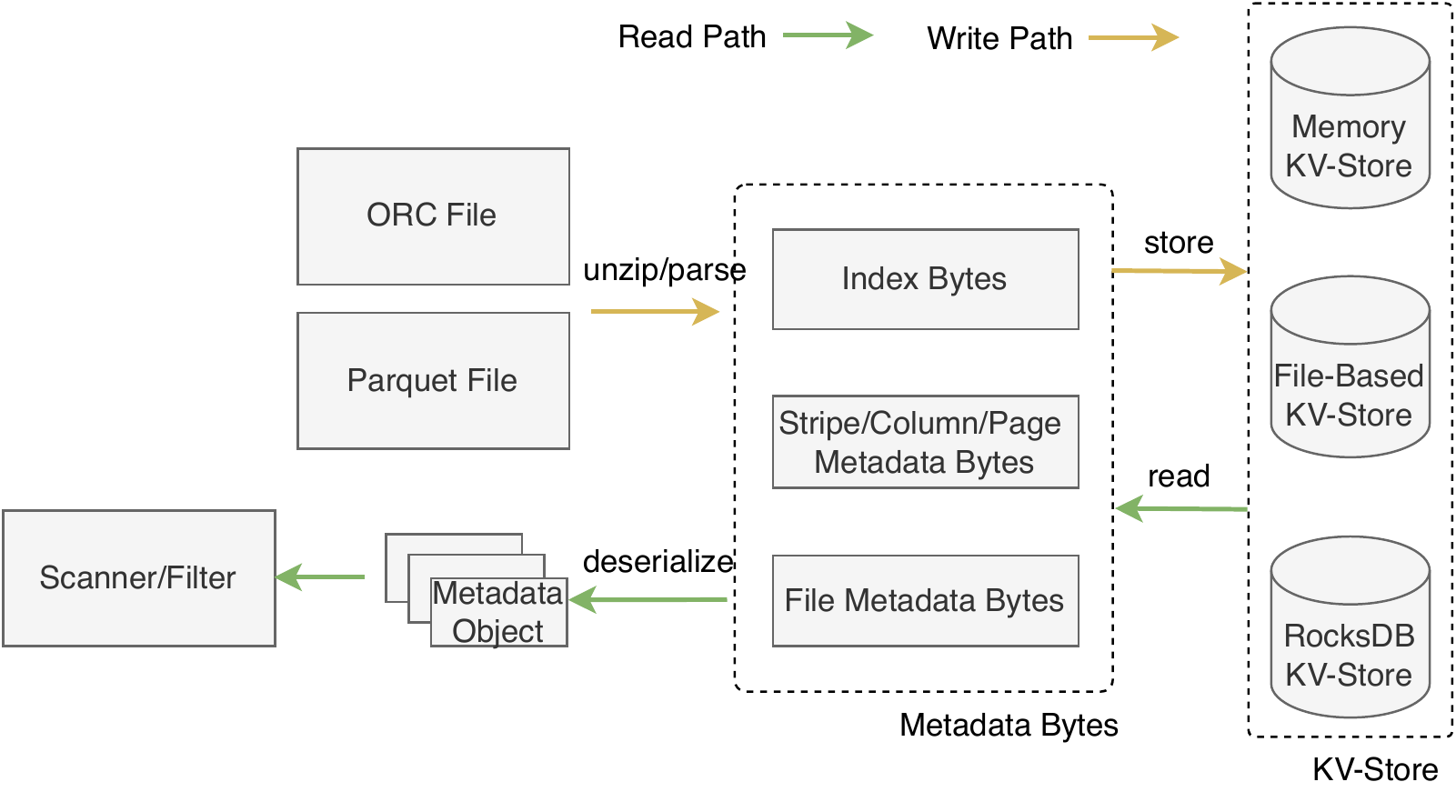}}
    \caption{High-level design for caching decompressed metadata bytes.}
    \label{fig.strategy-1}
\end{figure}

The first method involves caching decompressed metadata bytes from raw columnar files, including ORC and Parquet. Our metadata cache is aware of the concrete file format and handles them separately. Figure \ref{fig.strategy-1} explains the details of the design of the write path and the read path.

When a Presto worker node reads a file stored in a column-oriented format, it parses and decompresses the file to obtain index, stripe/column/page metadata, and file metadata. Then the worker node writes the extracted metadata bytes to a key-value store. When a SQL query comes in and needs the cached objects, the Presto reader will load the bytes from the key-value store and deserialize them into in-memory metadata objects. Then the Presto worker runs scanning and filtering operations to complete the scheduled execution plan.

\begin{figure}[thp]
    \centerline{\includegraphics[width=0.45\textwidth]{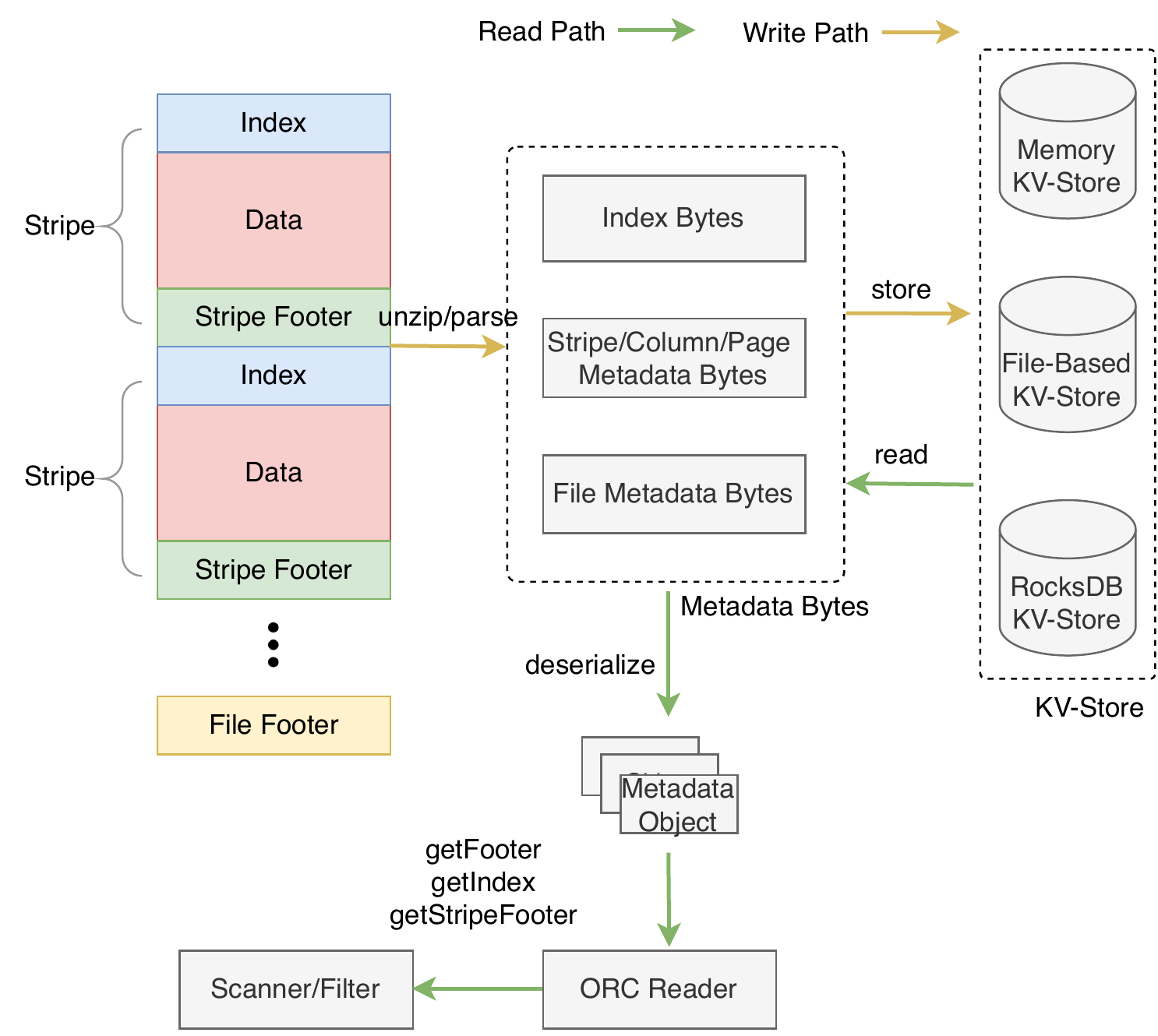}}
    \caption{Caching decompressed metadata bytes from ORC files.}
    \label{fig.strategy-1-orc}
\end{figure}

We demonstrate a concrete example of handling ORC files in Figure \ref{fig.strategy-1-orc}. ORC files are organized into groups of rows called stripes. Each stripe contains an index, data, and stripe footer. The index stores statistical values of this stripe, including max and min values, and row positions for each column. The stripe footer stores the directory of stream (column chunk) locations. There is also a file footer for each ORC file. The footer records the metadata of the file, such as the list of stripes and the number of rows in each stripe. Following the high-level caching method design in Figure \ref{fig.strategy-1}, the Presto ORC reader decompresses the ORC file, extracts the metadata, and writes the metadata bytes to key-value stores. And when the cached bytes are read, it loads the cached bytes from the store, deserializes them into in-memory objects, and calls the \textit{getFooter}, \textit{getIndex}, and \textit{getStripeFooter} functions for further processing.

\subsection{Method II - Caching Deserialized Metadata Objects}

\begin{figure}[thp]
    \centerline{\includegraphics[width=0.48\textwidth]{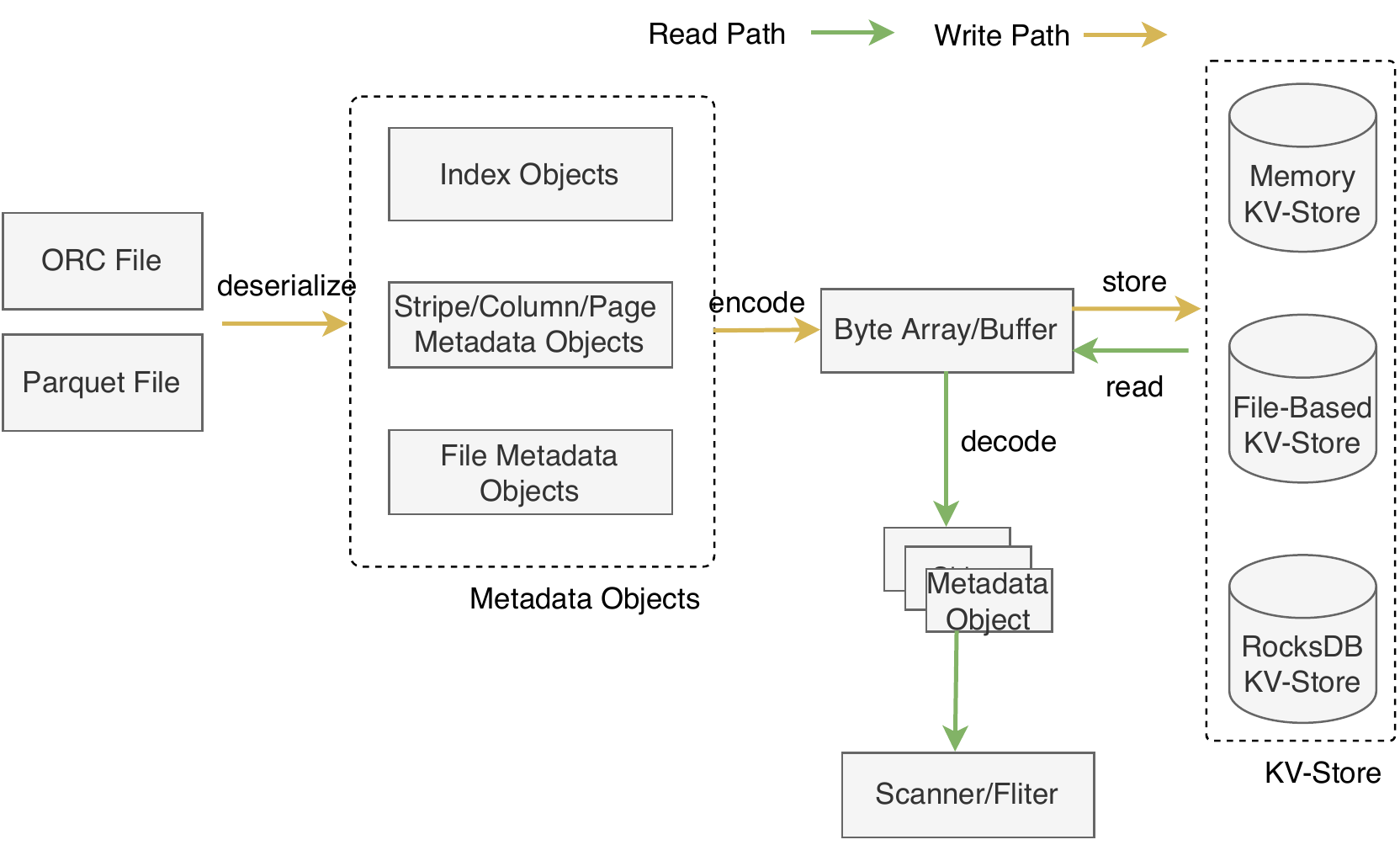}}
    \caption{High-level design for caching deserialized metadata objects.}
    \label{fig.strategy-2}
\end{figure}

The former method has a high write performance but a potentially downgraded read performance when cached objects are frequently accessed because each read operation requires a deserialization step. This deserialization step helps generate metadata objects from metadata bytes before scanning and filtering operations. To improve the read performance, we implement the second method by caching the deserialized metadata objects in byte buffers, as shown in Figure \ref{fig.strategy-2}.

In contrast to the former approach, after the Presto reader extracts metadata information from raw files, we deserialize the metadata bytes into metadata objects and encode them into byte buffers. The byte buffers have multiple choices and we choose Flatbuffers thanks to their zero-copy nature and high encoding/decoding performance. So when we read cached byte buffers from the key-value store, we only need to decode them into in-memory metadata objects, which is much more efficient than deserializing raw metadata bytes (Method I). This approach has a higher read performance for frequently accessed data in the cache with a minimal sacrifice of the write performance. Our evaluation in the next section explains our findings in detail.

\begin{figure}[thp]
    \centerline{\includegraphics[width=0.48\textwidth]{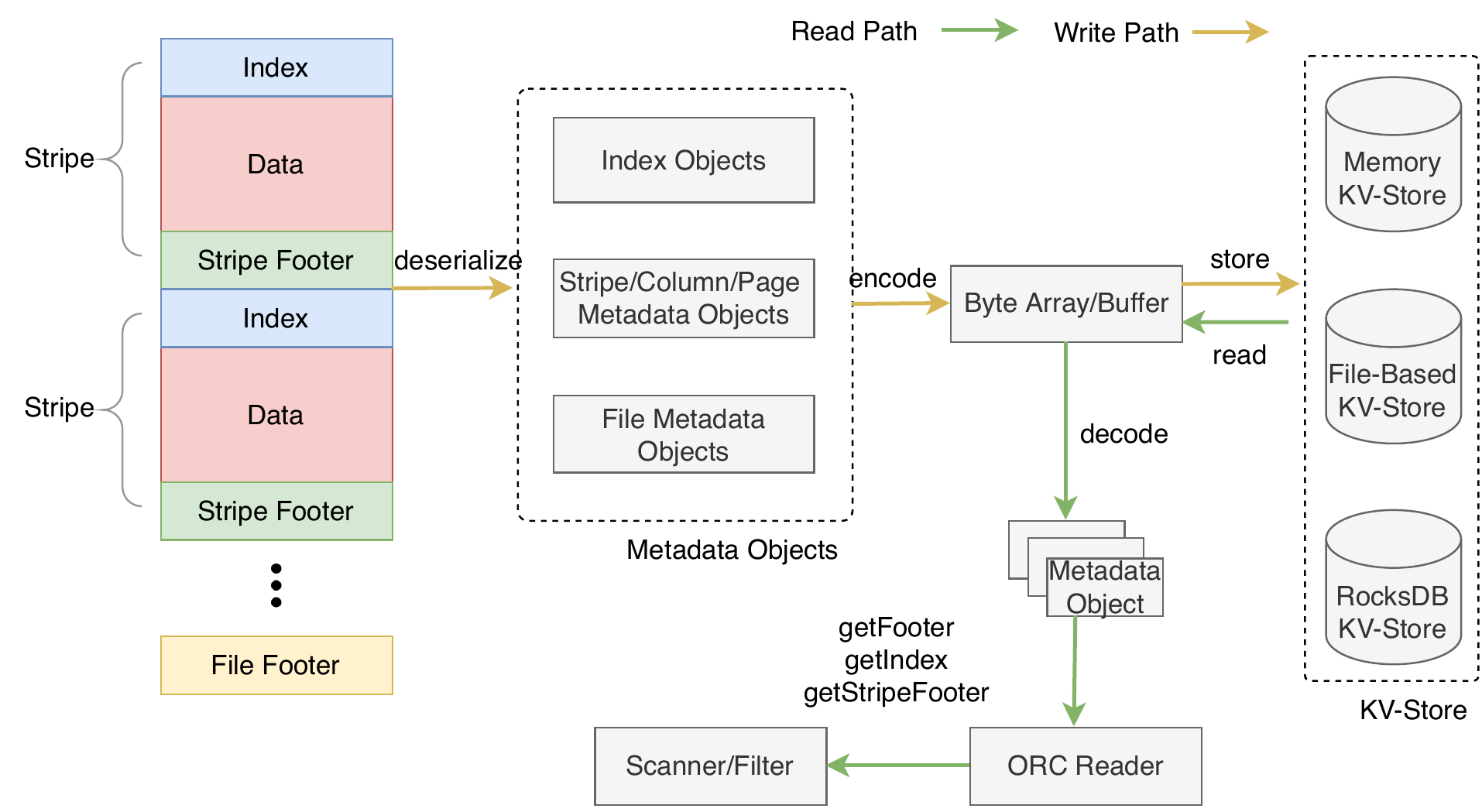}}
    \caption{Caching deserialized metadata objects from ORC files.}
    \label{fig.strategy-2-orc}
\end{figure}

Similar to the ORC example for the former method, we also show an ORC example for caching deserialized metadata objects in Figure \ref{fig.strategy-2-orc}. As we described, the ORC file's metadata is encoded into byte buffers and stored in key-value stores. When we read the metadata from the store, we decode the buffer into objects and feed them to Presto readers.

\section{Experimental Evaluation}
\label{sec.evaluation}

We investigate the effectiveness of our metadata cache by running the first 10 queries of the TPC-DS queries \cite{presto-tpcds} on Presto. These queries cover the typical scenarios of dashboard reporting and interactive analytics. We set up the experiments by deploying a Presto cluster with 1 coordinator node and 5 worker nodes. The benchmarked dataset is stored in an HDFS cluster with 5 datanodes in an on-premises Kubernetes cluster. As the bottleneck of computation reported is extremely high CPU consumption, we monitor, collect, and report metrics related to CPU usage, such as CPU time. We don't collect wall clock times as a query's computation tasks are executed in parallel in Presto, preventing the wall clock time from being a reliable metric for CPU consumption. We evaluate the cache read and write performance of three methods: No cache (baseline), Method I, and Method II. 

\begin{figure}[thp]
    \centerline{\includegraphics[width=0.48\textwidth]{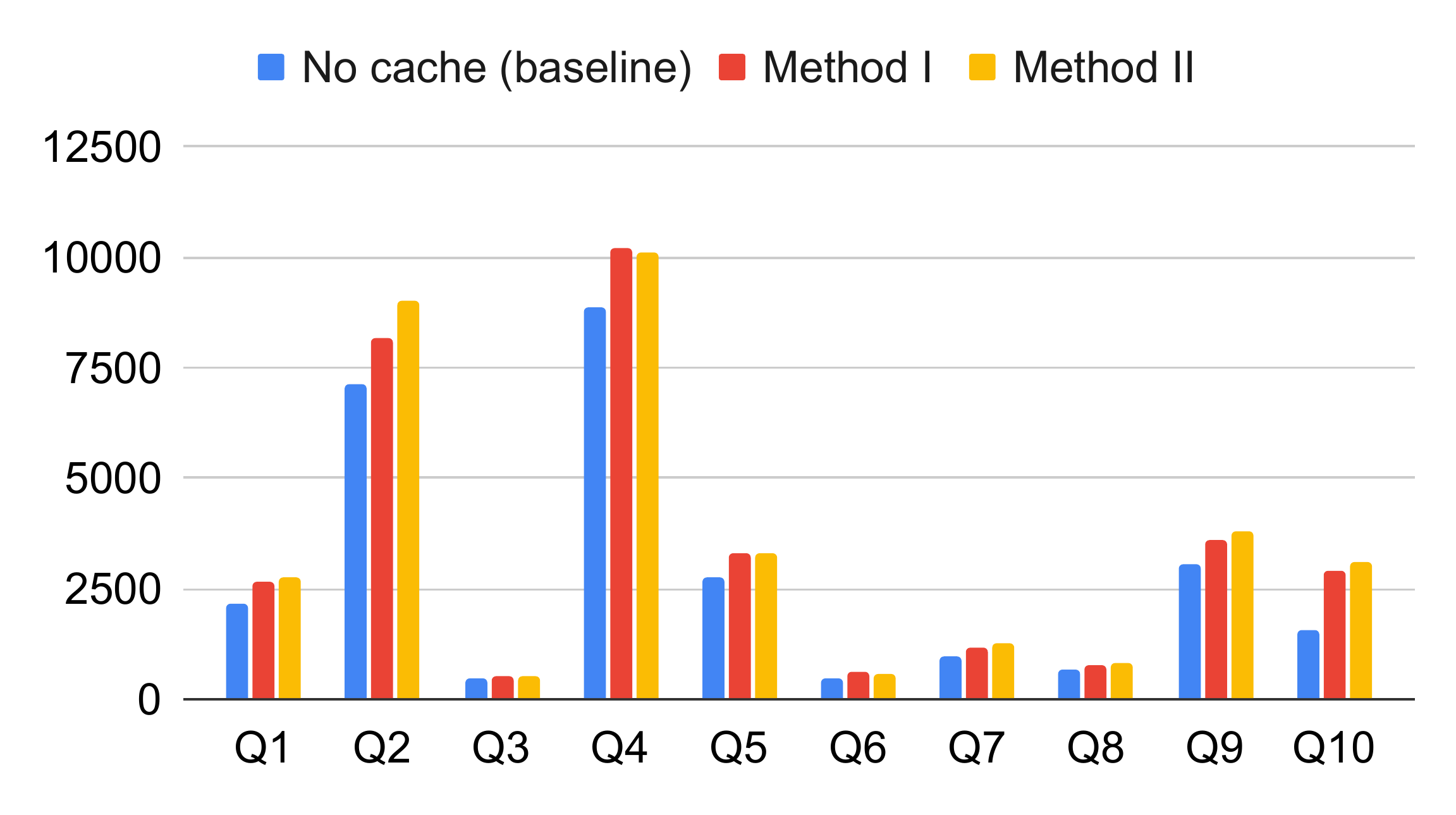}}
    \caption{Cache write performance of the three methods: No cache, Method I, and Method II. Evaluated in the total CPU time (ms) consumption for processing Query 1-10 (lower is better).}
    \label{fig.write-path-eval}
\end{figure}

To trigger a cache write, we initiate our experiments with a cold cache, so no metadata information is stored in the cache beforehand. An incoming SQL query will cause a cache miss and lead to a cache write operation. We report the total CPU time (ms) consumption of the three methods in Figure \ref{fig.write-path-eval}. From the figure, we observe that both our metadata caching methods bring performance overhead, which is expected as we need extra steps to cache the intermediate processing results. Method II also has a bit higher performance overhead than Method I, because it needs an extra encoding step before writing the metadata information to key-value stores. Method I's performance overhead is around 10-20\%, compared with the baseline; Method II's performance penalty ranges from 10\% to 30\%, which is a bit higher than Method I's. The only exception to the overhead ranges of Method I and II is Query 10, which involves computation across 6 tables. Under that scenario, both the two methods need to write more intermediate results to the metadata cache, leading to abnormal performance overhead.

\begin{figure}[thp]
    \centerline{\includegraphics[width=0.48\textwidth]{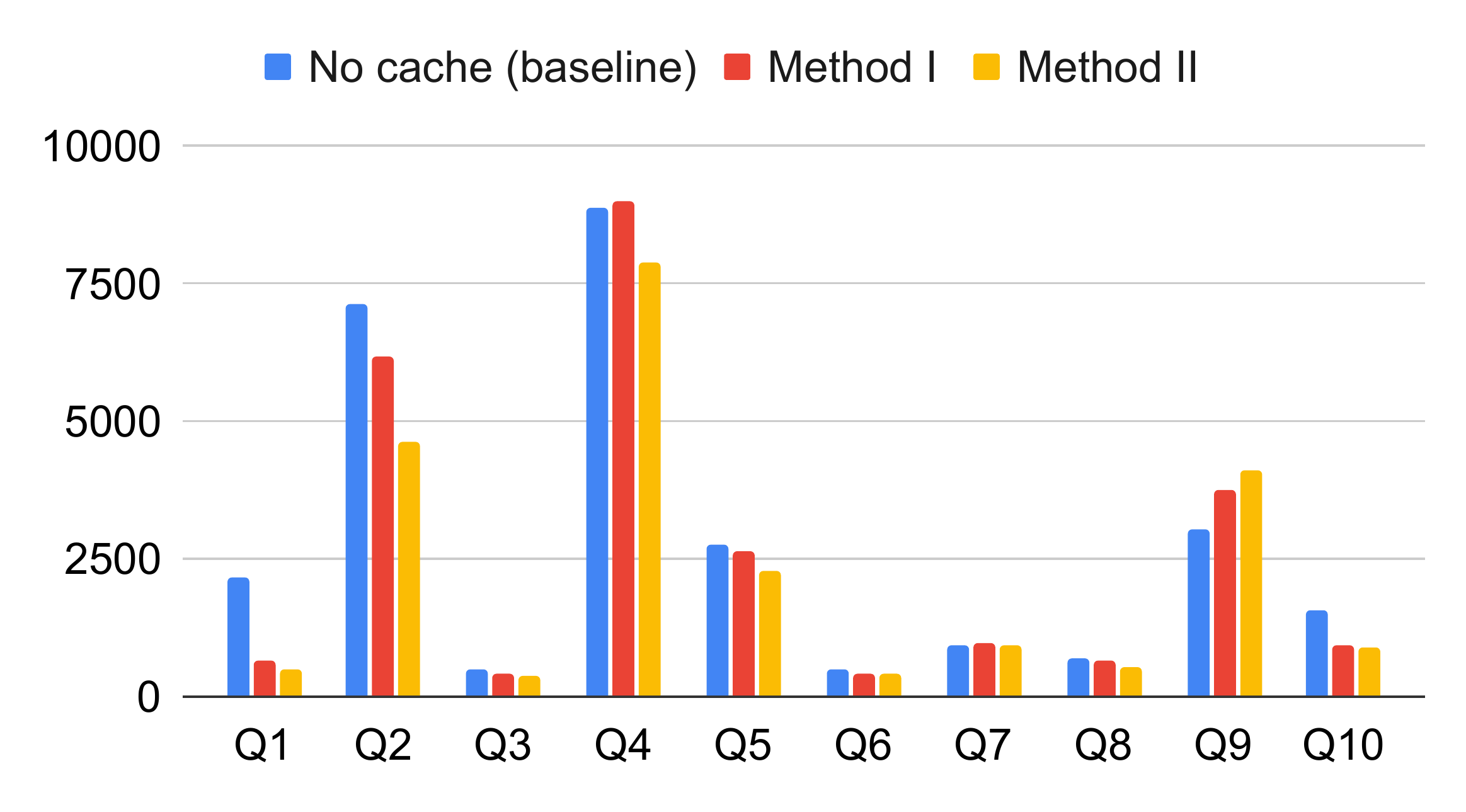}}
    \caption{Cache read performance of the three methods: No cache, Method I, and Method II. Evaluated in the total CPU time (ms) consumption for processing Query 1-10 (lower is better).}
    \label{fig.read-path-eval}
\end{figure}

When metadata is cached, the following SQL queries will trigger cache read operations. We collect and report the performance gain of our two caching methods in Figure \ref{fig.read-path-eval}, evaluated in the total CPU time (ms) consumption of processing the queries. It is worth mentioning that as the baseline approach does not employ any cache, the query performance reported in this figure is the same as that in Figure \ref{fig.write-path-eval}. We observe significant query performance improvement with the metadata cache: Method I reduces the CPU time consumption by around 10\%-20\%; Method II brings down the CPU time consumption by around 20\%-40\%. The two approaches especially work well on Query 1 which runs lots of table scan operations. They perform adversely on Query 9 (Method I worsens the performance by 23\% and Method II by 35\%). After investigation, we notice that the query requires more than 10 join operations, therefore the cache occupies a large chunk of memory, which consequently costs more CPU time on the Presto task scheduling.

\section{Related Work}
\label{sec.related-work}

The need for faster query processing has led to the adoption of cache in storage and computation of OLAP systems. Here, we discuss related work in this domain.

Some prior projects employ cache to store intermediate results in query processing. This design helps to improve query performance with reusable intermediate computation results. For example, Durner et al. \cite{durner2021crystal} developed a unified cache storage system, that is co-located with compute nodes in analytical databases. They reported significantly improved query latencies and network bandwidth consumption by caching intermediate results of selected database operations. Presto supports Alluxio cache service \cite{presto-alluxio} to cache intermediate data. Presto community reported \cite{raptorx} the hierarchical caching design, including metastore version cache, file list cache, fragment result cache, can make Presto run as much as 10 times faster. Various cache policies are also discussed and studied for caching the results. For instance, Azim et al. \cite{azim2017recache} presented a reactive cache-based performance accelerator on heterogeneous data across various raw file formats, with dynamic cache admission and eviction policies. Yang et al. \cite{yang2021flexpushdowndb} combined caching and computation pushdown in OLAP cloud databases with a novel cache-replacement policy, which outperforms the baseline LFU by 37\%. 

Materialized view \cite{gupta1995maintenance} is another popular technique using cache in database optimization. It caches the result of a query in a separate table. Some recent work includes automated material view generation \cite{ahmed2020automated} and optimized view selection \cite{gosain2019selection}. But as building and maintaining material view is usually time-consuming and resource-consuming, we do not adopt material view in our cache design.

\section{conclusion}
\label{sec.conclusion}

In this paper, we address the need for reducing CPU consumption for parsing column-oriented data files by introducing a metadata caching mechanism in Presto worker nodes. The metadata cache can store either the decompressed metadata bytes or the deserialized metadata objects. Our evaluation results show that the two caching methods, in particular the caching deserialized metadata objects approach, can significantly improve the CPU consumption associated with parsing data files, with minimal performance overhead during a cold start.

\section*{Acknowledgment}

Both Presto and Alluxio communities have actively evolved for years, with the help of a large number of contributors. We would like to express our gratitude to Tim Meehan, Nezih Yigitbasi, Andrii Rosa, James Petty, James Sun, Jiexi Lin, Leiqing Cai, Maria Basmanova,  Nikhil Collooru, Rebecca Schlussel, Shixuan Fan, Venki Korukanti, Wenlei Xie, Ying Su, Zhenxiao Luo, and many others from the Presto community; Haoyuan Li, Amelia Wong, Xiaodan (Danica) Wang, Sridhar Venkatesh, Adit Madan, Lu Qiu, David Zhu, Zhengjia Fu, Jiacheng Liu, Yimin Wei, Yong Yang, Jianjian Xie, Shawn Sun, Jingwen Ouyang, Arthur Jenoudet, Rico Chiu, Saiguang Che, and many others from Alluxio. We also thank the anonymous SCDM reviewers for their informative comments, which considerably improved our paper.

\bibliographystyle{IEEEtran}
\bibliography{library}

\end{document}